\documentclass[conference]{IEEEtran}

\usepackage[noadjust]{cite}
\usepackage{blindtext} 
\usepackage{amsmath} 
\usepackage{amsfonts} 
\usepackage{graphicx} 
\usepackage{tabulary}
\usepackage{xcolor} 
\usepackage{pifont}
\newcommand{\cmark}{\ding{51}}
\newcommand{\xmark}{\ding{55}}

\ifCLASSOPTIONcompsoc 
\usepackage[caption=false,font=normalsize,labelfon
t=sf,textfont=sf]{subfig}
\else
\usepackage[caption=false,font=footnotesize]{subfig}
\fi

\newcommand{\tsup}{\textsuperscript}
\newcommand{\tsub}{\textsubscript}
\DeclareMathOperator*{\argmin}{arg\,min} 
\newcommand{\rom}[1]{\uppercase\expandafter{\romannumeral #1\relax}}

\begin{document}
	\bstctlcite{BSTcontrol} 
	
	\title{Low Complexity Channel Model for Mobility Investigations in 5G Networks}
	
	\author{\IEEEauthorblockN{
			Umur Karabulut\IEEEauthorrefmark{1}\IEEEauthorrefmark{2}, 
			Ahmad Awada\IEEEauthorrefmark{1}, 
			Ingo Viering\IEEEauthorrefmark{3}, 
			Andre Noll Barreto\IEEEauthorrefmark{4} and 
			Gerhard P. Fettweis\IEEEauthorrefmark{2}}
		\IEEEauthorblockA{
			\IEEEauthorrefmark{1}Nokia Bell Labs, Munich, Germany, 
			\IEEEauthorrefmark{2}Vodafone Chair Mobile Communications Systems, Technische Universit\"at Dresden\\
			\IEEEauthorrefmark{3}Nomor Research GmbH, Munich, Germany, 
			\IEEEauthorrefmark{4}Barkhausen Institut gGmbH, Dresden, Germany
		}
	}
	
	\maketitle
	
	\begin{abstract}
	Millimeter-wave has become an integral part of 5G networks to meet the ever-increasing demand for user data throughput. Employing higher carrier frequencies introduces new challenges for the propagation channel such as higher path loss and rapid signal degradations. On the other hand, higher frequencies allow deployment of small-sized antenna elements that enable beamforming. To investigate user mobility under these new propagation conditions, a proper model is needed that captures spatial and temporal characteristics of the channel in beamformed networks. Current channel models that have been developed for 5G networks are computationally inefficient and lead to infeasible simulation time for most user mobility simulations. In this paper, we present a simplified channel model that captures the spatial and temporal characteristics of the 5G propagation channel and runs in feasible simulation time. To this end, coherence time and path diversity originating from fully fledged Geometry based Stochastic Channel Model (GSCM) are analyzed and adopted in Jake's channel model. Furthermore, the deviation of multipath beamforming gain from single ray beamforming gain is analyzed and a regression curve is obtained to be used in the system-level simulations. We show through simulations that the proposed simplified channel model leads to mobility results comparable to Jake’s model for high path diversity. Moreover, the multi-path beamforming gain increases the interference in the system and in turn number of mobility failures.
	\end{abstract}
	

	\section{Introduction}
		
	The available spectrum in lower frequency ranges does not meet the unprecedented increase in demand for user data throughput in mobile networks. Third Generation Partnership Project (3GPP) thus designated for Fifth Generation (5G) mobile network technology the use of millimeter-wave (mmWave) frequency bands along with frequencies below 6 GHz, due to the availability of larger transmission bandwidth.
	
	Enabling higher carrier frequencies introduces new channel conditions. Propagating signals are exposed to higher diffraction loss \cite{DiffLoss} and are highly susceptible to blockage caused by surrounding objects, which leads to rapid signal degradations \cite{bodyblockage}. On the other hand, higher carrier frequencies enable the deployment of many small-sized antennas that are used for directional signal transmission, resulting in beamforming gain. 
	
	Higher diffraction loss and rapid signal degradation caused by obstructions pose a challenge for user mobility at mmWave frequencies. Besides, measurements used for user mobility and handover decisions are impaired by fluctuations that are caused by measurement error and fast-fading. The spatial and temporal characteristics of fast-fading are highly correlated with carrier frequency. Mobility investigations require long simulation time to collect handover and failure statistics. For better understanding of user mobility at higher carrier frequencies, an efficient simulator is needed that runs in feasible simulation time and captures the spatial and temporal characteristics of the mmWave channel which are relevant for user mobility investigations \cite{Commag}.
	
	There are numerous studies \cite{QuaDRiGa,winner2,M2Mmodel,38901,NYUSIM} that model the radio propagation channel for 5G based on measurements for both lower and higher carrier frequencies. The Quasi Deterministic Radio Channel Generator (QuaDRiGa) \cite{QuaDRiGa} extends the Wireless Initiative for New Radio (WINNER) \cite{winner2} model with 3D propagation and antenna patterns along with time evolution and scenario transitions between Line-of-Sight (LOS) and Non-LOS (NLOS). In \cite{M2Mmodel}, the authors address a channel model for indoor Machine-to-Machine (M2M) applications. The 3rd Generation Partnership Project (3GPP) and New York University (NYU) have also developed 3D channel models \cite{38901,NYUSIM}, based on real-world channel measurements at various frequencies (including mmWave) for different scenarios such as urban microcell (UMi), urban macrocell (UMa), and rural macrocell (RMa). 
	
	 The channel models mentioned above can be good candidates for mobility simulations.However, they are computationally inefficient, which leads to infeasible simulation time \cite{Commag}. In this paper, a simplified channel model is developed based on Geometry-based Stochastic Channel Model (GSCM) \cite{38901}, which runs in feasible simulation time and captures the spatial and temporal characteristics of \cite{38901} for different receiver (RX) speeds, carrier frequencies, LOS and NLOS conditions. Besides, the proposed channel model is designated for beamformed systems, where directional transmission is enabled by deployment of multiple antenna elements. Moreover, the impact of this simplified channel model and Jake's channel model on user mobility performance are compared in a 5G mmWave scenario.
	
	This paper is organized as follows. Jake's fading model \cite{Jakes} and GSCM \cite{38901} are reviewed in Section~\ref{sec:ChannelModels}. Our simplified channel model proposed for New Radio (NR) user mobility investigations is presented in Section~\ref{sec:AbstractChannelModel}. The simulation models and scenario are explained in Section~\ref{sec:SimScenario}. Simulation results are presented in Section~\ref{sec:PerformanceEvaluation} to compare the proposed channel model against Jake's channel model with respect to mobility performance. The paper is then concluded in Section~\ref{sec:Conclusion}.
		
	\section{Channel Models}
	\label{sec:ChannelModels}
	In this section, the well-known Jake's fading model \cite{Jakes} is reviewed first. Then, observations of spatial and temporal characteristics of GSCM \cite{38901}, leading to the proposed simplified model, are presented.
	
	\subsection{Jake's Fading}
	In wireless channels, multiple replicas of the transmitted wave (multipath waves) arrive at the RX from all directions since the wave is exposed to diffraction, reflection and scattering caused by surrounding objects in the wireless medium which is known as multipath propagation. Due to the multipath propagation, multipath waves are phase-shifted, delayed and attenuated replicas of emitted wave which arrive from all directions at the RX. Superposition of the waves at the RX can be either constructive or destructive, depending on the phase shift of each multipath waves which causes temporal variation (fluctuation) of the received signal power. 
	
	In addition to signal distortion caused by multipath propagation, the Doppler effect also has an adverse impact on the received signal for non-stationary users.. Let us assume a mobile RX, with a velocity $v$, which receives a wave from an angle $\theta$ between the direction of the RX's motion and the direction of the arrival of the wave. Due to the mobility of the RX unit, the Doppler effect causes a frequency shift $f_d$ of the incident wave which is called Doppler shift and is expressed as
	\begin{equation}
		\label{eq:doppler_shift}
		f_d = f_\textrm{max} cos(\theta).
	\end{equation}
	Here, $f_\textrm{max}$ is the maximum frequency shift proportional to the RX velocity $v$ and carrier frequency $f_\textrm{c}$ which is given as
	\begin{equation}
		\label{eq:max_doppler_shift}
		f_\textrm{max} = \frac{v}{c}f_\textrm{c},
	\end{equation}
	where $c$ is speed of light. Considering the fact that there are multiple incident waves from all directions, the carrier frequency expands from $f_\textrm{c}-f_\textrm{max}$ ($\theta=\pi $) to $f_\textrm{c}+f_\textrm{max}$ ($\theta=0 $) which is called Doppler spread.

	One of the most popular channel model is Jake's fading model that captures the multipath and Doppler characteristics of the channel \cite{Jakes}. It is assumed that the incident waves arrive in two-dimensional (horizontal) plane with equal power. Besides, it is also assumed that the angle of arrival $\theta_k$  and phase shift $\phi_k$ of each incident wave $k$ are uniformly distributed in the interval $[-\pi,\pi]$, and the power radiation pattern of the RX antenna is omni-directional. Hence, the channel impulse response $h(t)$ can be represented by the superposition of number $K$ of orthogonal sinusoids as
	\begin{equation}
		\label{eq:jake_h}
		h(t) = \lim_{K\to\infty}\frac{1}{\sqrt{K}}\sum_{k=1}^{K}e^{j(2\pi f_\textrm{max}\cos(\theta_k)t+\phi_k)}.
	\end{equation}
	The channel impulse response given in (\ref{eq:jake_h}) is a complex Gaussian random process with power envelope $P(t) = |h(t)|^2$  following non-central Chi-square distribution ($nc-\chi^2$) with $2$ degrees of freedom, unit mean and variance \cite{Proakis}. 
	
	In mobility studies, handover decision mechanisms and measurements are based on signal power rather than on the complex impulse response. The power envelope of Jake's model represents a fading channel with single path and paths can be maximum ratio combined 
	to obtain a power envelope for higher path diversity \cite{PathDiversity}. By using (\ref{eq:jake_h}), one can formulate the power envelope $P_L(t)$ of a channel with L-path diversity as
	\begin{equation}
	\label{eq:jakes_multipath}
		P_L(t) = \lim_{K\to\infty}\frac{1}{KL}\sum_{l=1}^{L}\left|\sum_{k=1}^{K}e^{j(2\pi f_{max}\cos(\theta_{k,l})t+\phi_{k,l})}\right|^2,
	\end{equation}
	which is $nc-\chi^2$ distributed with $2L$ degrees of freedom, unit mean due to the normalization and $\sigma^2_L=1/L$ variance \cite{Proakis}. Hence, the path diversity of the power envelope $P_L(t)$ is expressed as
	\begin{equation}
	\label{eq:path_div}
		L=\frac{1}{{\sigma^2_L}}.
	\end{equation}
	
	The statistical measure of the channel fluctuations caused by Doppler spread can be captured by the coherence time $T_c$ of the channel impulse response which is a representation of the correlation of fading channel samples over time. From \cite{Jakes} we know that the autocorrelation $R(\Delta t)$ of the power envelope of (\ref{eq:jake_h}) can be expressed as 
	\begin{equation}
		\label{eq:bessel}
		R(\Delta t) = {J_0}^2(2\pi f_{max} \Delta t),
	\end{equation}
	where $J_0$ is zero-order Bessel function of the first kind and $\Delta t$ is the scale of the time distance transversed for correlation. The time correlation in (\ref{eq:bessel}) also applies to (\ref{eq:jakes_multipath}), since temporal characteristics of the fading process do not change for varying path diversity. In practice, the coherence time $T_c$ is often defined as $50\%$ correlation of the $R(\Delta t)$ which can be approximated as
	\begin{equation}
		\label{eq:Tc}
		T_c \approx \frac{9}{16\pi f_\textrm{max}}.
	\end{equation}

	\subsection{Geometry-based Stochastic Channel Model (GSCM)}
	
	In 3GPP, GSCM is proposed \cite{38901} to properly study and evaluate the performance of upcoming 5G features.The GSCM is made up of stochastic and deterministic aspects of the channel where the former aspect reflects the spatial consistency based on the positions of scattering waves (clusters) and the latter represents the characteristics of the propagation environment. The statistical characteristics of the environment are captured by a large-scale fading (LSF) model where the joint spatial correlation of the LSF components are incorporated in the model. 
	
	The channel impulse response matrix between each antenna element of transmitter (TX) and RX is denoted as $\boldsymbol{H}(\tau,t)$ for a given cluster delay $\tau$ and time $t$ \cite{38901}. The number of TX and RX antenna elements is given by $S$ and $U$, respectively, and each element $h_{u,s}(\tau,t)$ of $\boldsymbol{H}(\tau,t)$ from column $u\in [1,U]$, and from row $s\in[1,S]$, $u,s\in\mathbb{N^+}$ represents the channel impulse response from TX antenna element $u$ to RX antenna element $s$. Modeling the channel impulse response $\boldsymbol{H}(\tau,t)$ is described for both line-of-sight (LOS) and non-LOS (NLOS) scenarios in \cite{38901}.
	
	Proper use of multiple antenna elements enables the directional signal transmission, also known as beamforming. The phase of the transmitted signal from each antenna element can be adjusted such that all signals add constructively in a specific direction, resulting in beamforming gain. For a given beam direction in 3-D plane elevation angle $\theta_b$, azimuth angle $\phi_b$ with beam direction index $b$, the phase shift $w_{s,b}$ of antenna element $s$ is formulated as
	
	\begin{subequations}
		\begin{align}
		w_{s,b} =& e^{j2\pi \vec{\boldsymbol{r}}_b^T \vec{\boldsymbol{d}}_s},\\
		\vec{\boldsymbol{r}}_b =& \begin{bmatrix} \sin(\theta_b)\cos(\phi_b) \\ \sin(\theta_b)\sin(\phi_b) \\ \cos(\theta_b) \end{bmatrix}, \\
		\vec{\boldsymbol{d}}_s =& \begin{bmatrix} x_s\\y_s\\z_s \end{bmatrix}.
		\end{align}
	\end{subequations}
	Here, $\vec{\boldsymbol{r}}_b$ is the unit vector that is directed towards the beam direction and $\vec{\boldsymbol{d}}_s$ holds the 3D coordinates of transmit antenna element $s$ in Cartesian coordinates. In order to apply beamforming to the channel impulse response matrix $\boldsymbol{H}(\tau,t)$, the weighted sum of the channel impulse response for each transmit antenna element $s$ is evaluated as
	\begin{equation}
	\label{eq:beam_impulse}
			h^B_{b,u}(\tau,t) = \frac{1}{\sqrt{S}}\sum_{s = 1}^{S} w_{s,b} h_{u,s}(\tau,t).
	\end{equation}
	where $h^B_{b,u}(\tau,t)$ is the channel impulse response of beam $b$ from TX to RX antenna element $u$ is obtained in (\ref{eq:beam_impulse}).
	
	In case of beamforming, cluster power does not follow the uniform distribution at the RX due to the directional transmission, hich increases cluster power towards the intended direction and suppresses it in unintended directions. As such, the Doppler spread of the channel impulse decreases which leads to an increase in the coherence time of the channel impulse response as expressed in (\ref{eq:Tc}). 
	
	\begin{figure}[!htb]
		\centering
		\subfloat[Coherence time distribution per beam.] {\includegraphics[width=0.95\columnwidth]{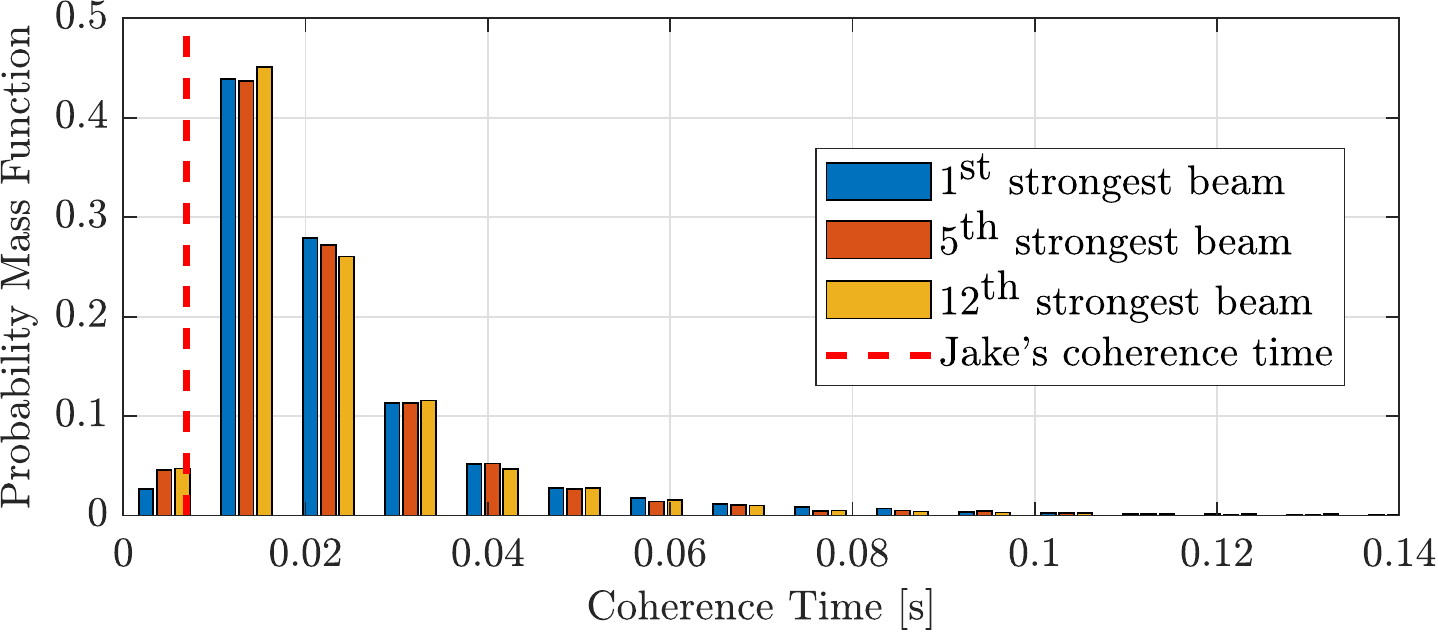}\label{fig:PerBeamTc}}
		\hfil
		\subfloat[Path diversity distribution per beam where $ 20+ $ indicates path diversity value of 20 and above.] {\includegraphics[width=0.95\columnwidth]{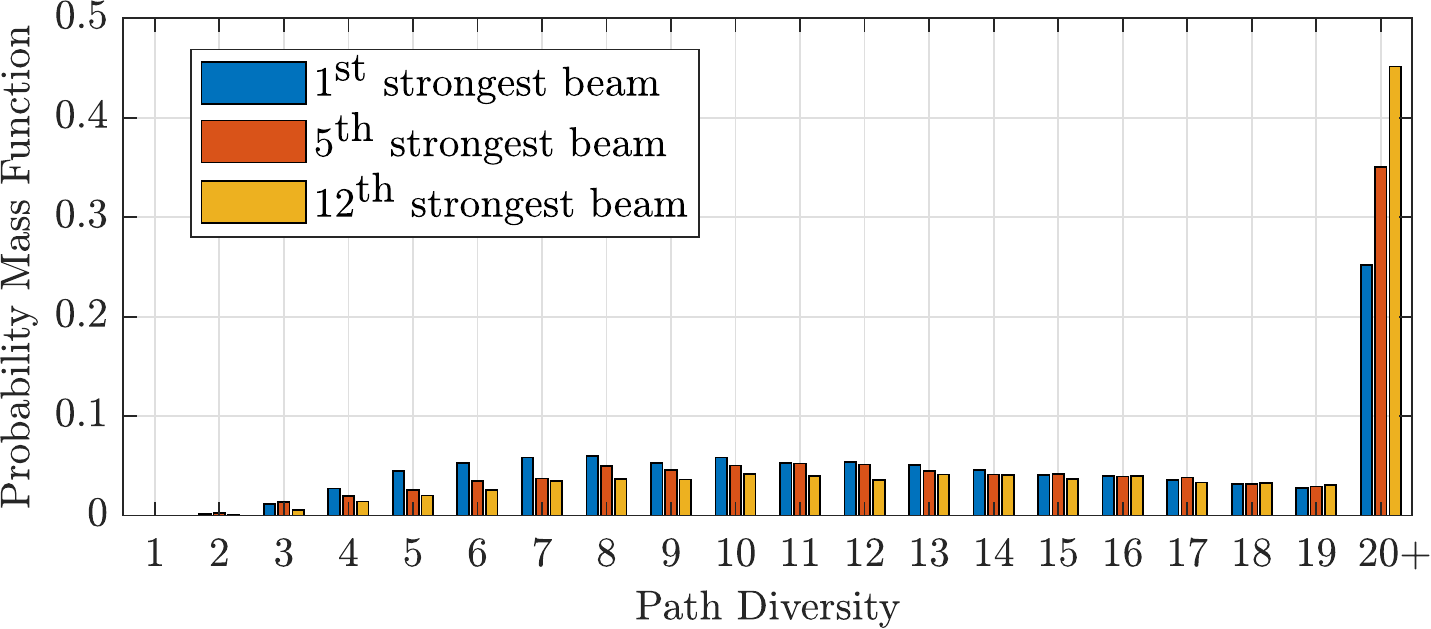}\label{fig:PerBeamDoF}}
		\caption{Coherence time and path diversity distribution of 1\tsup{st}, 5\tsup{th} and 1\tsup{th} strongest beams for GSCM.}
		\label{fig:PerBeamStats}
	\end{figure}

	Figure \ref{fig:PerBeamStats} shows the spatial and temporal statistics of the power envelope of GSCM NLOS case for the parameter configuration given in Table~\ref{tab:sim_parameters1}. There are 12 beam directions, where beam elevation angles $\theta_b$ are defined in (\ref{eq:beam_ele}), beam azimuth angles $\phi_b$ and antenna panel sizes $\Omega_b$ are given in Table~\ref{tab:sim_parameters1}. Channel impulse response $ h_{u,b}^B(\tau,t)$ (\ref{eq:beam_impulse}) is evaluated for beams $ b\in[1,12] $. Then, the empirical probability mass function of coherence time $T_c$ and path diversity $L$ of 1\tsup{st}, 5\tsup{th} and 12\tsup{th} strongest beams are illustrated in Figure~\ref{fig:PerBeamTc} and Figure~\ref{fig:PerBeamDoF} respectively. The strength of the beam is evaluated by the average power of impulse response over time. The coherence time of each process is estimated by $50\%$ auto-correlation of the power envelope, and $50\%$ coherence time for Jake's model shown in (\ref{eq:Tc}) is given by the red dashed line as reference. It is assumed that the power envelope of the channel impulse response in (\ref{eq:beam_impulse}) follows $nc-\chi^2$ distribution with unit mean and $\sigma^2_L=1/L$ variance. Hence, the path diversity of each process is approximated and quantized to closest integer value by using the statistical property of the distribution given in (\ref{eq:path_div}).

	\begin{equation}
	\label{eq:beam_ele}
	\centering
	\theta_b =
	\begin{cases}
	-52.5+15(b-1) & b\in[1,8] \\
	-45+30(b-8) & b \in [9,12] \\
	\end{cases}
	\end{equation}

	\begin{table}[!htb]
		\renewcommand{\arraystretch}{1.3}
		\caption{Simulation Parameters \rom{1}}
		\label{tab:sim_parameters1}
		\centering
		\begin{tabulary}{\columnwidth}{L L}
			\hline 
			Parameter & Value \\ 
			\hline 
			\hline 
			Carrier frequency $f_\textrm{c}$ & 28 GHz \\ 
			
			RX velocity $v$ & 1 km/h \\ 
			
			System bandwidth & 100 MHz \\
			
			TX antenna height & 10m\\
			
			RX height & 1.5m\\
			
			TX panel size $\Omega_b$ & $16 \times 8,~\forall b\in [1,8]$; $8 \times 4,~\forall b \in [9,12]$\\
			Beam azimuth angle $\phi_b$ & $90,~\forall b \in [1,8]$; $97,~\forall b \in [9,12]$\\		
			TX Antenna element pattern & Table~7.3-1 in \cite{38901} \\

			TX vertical antenna element spacing & $0.7 \lambda$  \\
			
			TX horizontal antenna element spacing & $0.5 \lambda$  \\ 
			
			RX antenna element pattern & isotropic \\
			
			RX antenna element gain & 0 dBi \\
			
			Scenario & UMi-Street Canyon \cite{38901}\\
			
			\hline 
		\end{tabulary} 
	\end{table}

	The coherence time distribution given in Figure~\ref{fig:PerBeamTc} shows that $95\%$ of the coherence time of the beams are above Jake's coherence time due to beamforming. It is also shown that the temporal characteristics of the channel impulse response do not vary much among beams. However, Figure~\ref{fig:PerBeamDoF} shows that the path diversity is correlated with the strength of the beam, where the path diversity of the beam impulse response decreases for the beams with higher average power. 
	
	\begin{figure}[!htb]
	\centering
	\subfloat[LOS case.] {\includegraphics[width=0.95\columnwidth]{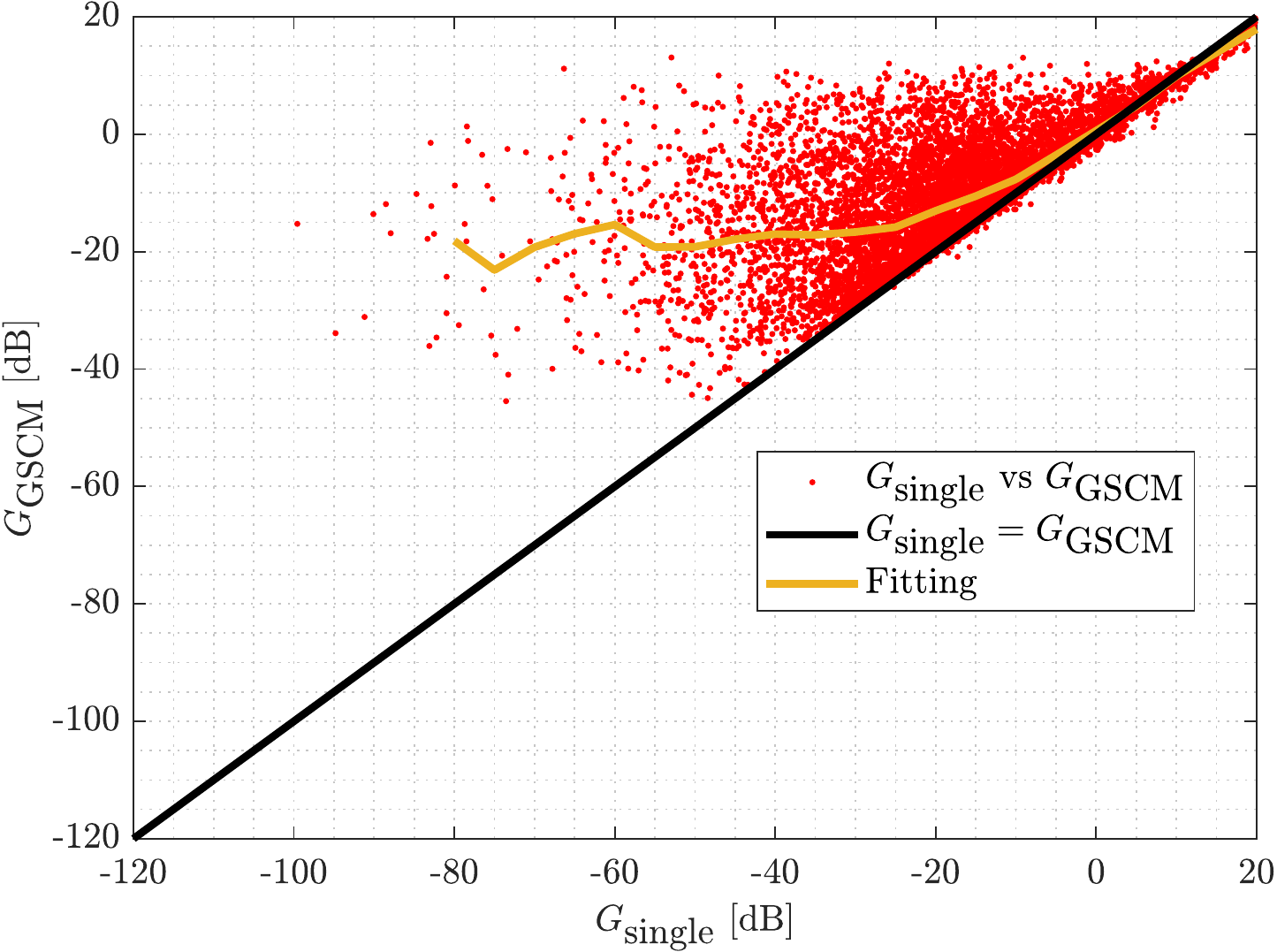}\label{fig:BFG_LOS}}
	\hfil
	\subfloat[NLOS case.] {\includegraphics[width=0.95\columnwidth]{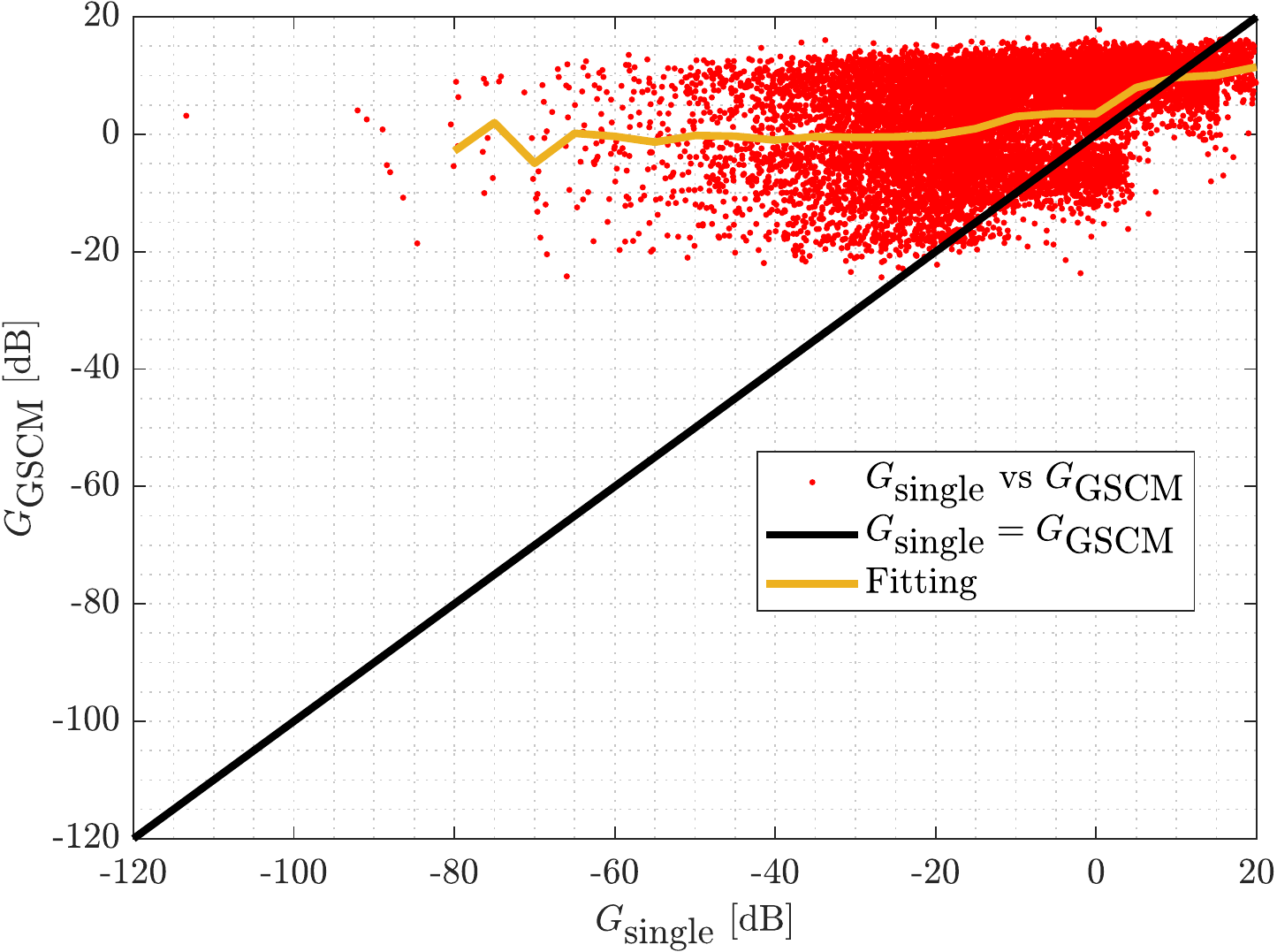}\label{fig:BFG_NLOS}}
	\caption{Beamforming gain deviation for LOS and NLOS cases. Beamforming gain of GSCM $ G_\textrm{GSCM} $ is evaluated for beamforming gain of single ray $ G_\textrm{single} $ (red points). Linear regression that is applied on beamforming gains is shown by fitting (orange line)}.
	\label{fig:BFG_all}
	\end{figure}
	
	In typical system-level simulators, it is assumed that there is a single ray between TX and RX in LOS direction $ \theta_{\textrm{LOS}} $ and $ \phi_{\textrm{LOS}} $. Eventually, the beamforming gain $ G_{\textrm{single}} $ of the link between TX and RX is calculated for a single ray in the LOS direction. Considering the fact that multipath propagation consists of multiple rays with different angle of departure and arrival, the beamforming gain of a multipath channel deviates in principle from $ G_{\textrm{single}} $. The beamforming gain of the multipath channel $ G_{\textrm{GSCM}} $ is modeled as the mean power of the channel $ h_{b,u}^B(\tau,t) $ normalized to the mean power of the channel impulse response $ h_{u,s} (\tau,t) $ for any $u$ and $s$.
	
	 $10^4$ realizations of $G_{\textrm{GSCM}}$ are simulated, and in each simulation $G_{\textrm{single}}$ is evaluated for each $G_{\textrm{GSCM}}$ to visualize the deviation of $G_{\textrm{GSCM}}$ from $G_{\textrm{single}}$. In Figure \ref{fig:BFG_all}, $G_\textrm{GSCM}$ is shown as a scatter plot (red) against $ G_{\textrm{single}} $ along with linear regression $ G_\textrm{fitting} $ shown in orange that is applied on the collected beamforming gains. The black line shows the reference where $G_{\textrm{GSCM}}$ does not deviate. For larger beamforming gains in the LOS case (Figure \ref{fig:BFG_LOS}), $G_{\textrm{GSCM}}$ does not deviate much from $G_{\textrm{single}}$. However, for decreasing $G_{\textrm{single}}$, $G_{\textrm{GSCM}}$ saturates around $-20$ dB, which is caused by the scattering structure of GSCM,\ where some of the clusters are not attenuated as much as the single ray in LOS direction due to the angular spread at TX. In the NLOS case (Figure \ref{fig:BFG_NLOS}), $G_{\textrm{single}}$ is larger than $G_{\textrm{GSCM}}$ for high beamforming gain values. This is because higher $G_{\textrm{single}}$ values are observed in case the transmission is on boresight direction of the beam where GSCM applies angular spread on transmit clusters and attenuates the clusters which are not on boresight direction. Similar to the LOS case, $G_{\textrm{GSCM}}$ saturates around $0$ dB against much smaller $G_{\textrm{single}}$ values.
	\section{Simplified Channel Model}
	\label{sec:AbstractChannelModel}
	\begin{figure*}[!t]
		\centering
		\includegraphics*[width=0.9\textwidth]{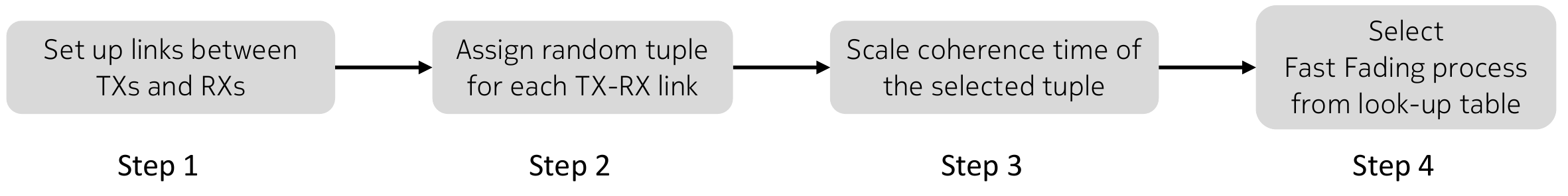}
		\caption{Steps of simplified model for fast-fading processes.}
		\label{fig:ff_process}
	\end{figure*}
	In system-level mobility simulations, a model for communication links between TX and RX is needed. In this section, a simplified fast-fading channel model is presented, which integrates the stochastic spatial and temporal characteristics of the GSCM into Jake's channel model. The evaluation of the fast-fading process for each link is composed of 4 steps as shown in Figure~\ref{fig:ff_process}. The description of each step is provided below.
	
	\subsubsection{Initial Simulation Setup}
	For a system level mobility simulation, it is assumed that the network topology and TX/RX locations are initialized. All possible combinations of the links $\forall e \in [1,E]$ between each TX and RX are evaluated, i.e. for a network with number $U$ of user equipments (UEs) as receiver and number $C$ of cells, there are $E = U\cdot C$ possible combinations of links between UEs and cells. 
	
	\subsubsection{Assign random tuple for each TX-RX link}
	Based on the network setup defined in Step 1, the parameters of the GSCM are configured such that the fast-fading process of GSCM is aligned with the statistics of the mobility simulation: $M$ fast-fading processes are generated for both LOS and NLOS cases, fixed carrier frequency $f_c$ and velocity $v$, and for each beam $b \in [1,B]$. Then, the path diversity $L(m,b)$  of process $m\in[1,M]$ is estimated by using the relation between path-diversity and variance of process given in (\ref{eq:path_div}). The coherence time $T_c(m,b)$ is also estimated by observing $50\%$ autocorrelation of the process as shown in (\ref{eq:Tc}). Estimated $T_c(m,b)$ and $L(m,b)$ of each process form a tuple where each tuple is $B\times 2$ matrix and each row contains $T_c(m,b)$ and $L(m,b)$ of beam $b$. Rows of the tuple matrix are sorted by mean power of each beam $b$ in descending order. Finally, for each link $e$, one LOS and one NLOS random tuple are drawn from the pre-processed tuple list.
	
	\subsubsection{Coherence Time Scaling}
	The randomly drawn coherence time $T_c(m,b)$ from the tuple list represents the statistics of a fixed carrier frequency $f_c$ and velocity $v$, which may not always match the simulated network setup configuration; i.e., in case a different value for $ f_c$ and $v$ are simulated. For a given carrier frequency $f_c'$ and velocity $v'$, a new coherence time $T_c'(m,b)$ is derived from the tuple coherence time $T_c(m,b)$ as follows
	\begin{equation}
		T_c'(m,b) = T_c(m,b)\frac{f_c v}{f_c' v'}.
		\label{eq:Tc_scale}
	\end{equation}
	\subsubsection{Selection of fast-fading process}
	For $I$ different path diversities $L_i, i\in [1,I]$ and $J$ different coherence times $T_{c,j}~j\in [1,J]$, power envelopes of the fast-fading process (\ref{eq:jakes_multipath}) are evaluated in pre-processing and stored in a fast-fading look-up table (LUT). The set of path diversity and coherence time in the LUT is determined such that it captures the distribution of path diversity and coherence time of GSCM for the given simulation setup (carrier frequency, antenna panel, bandwidth etc.) and UE velocities. 
	
	The fast-fading process of each link $e \in [1,E]$ is assigned by selecting the appropriate fast-fading process from the LUT. For any scaled coherence time $T_c'(k,b)$ of randomly drawn coherence time from a tuple and corresponding path diversity $L(m,b)$ (Step 2-3), the nearest smaller values of coherence time $T_{c,s}$ and path diversity $L_s$ are selected from the LUT as follows
	\begin{subequations}
	\begin{align}
	T_{c,s}=& \argmin_{T_{c,j}} \lvert T_c'(m,b) - T_{c,j} \rvert &,&\quad T_{c,j} < T_c'(m,b) \\
	L_s=& \argmin_{L_i} \lvert L(m,b) - L_i \rvert &,&\quad L_i < L(m,b). 
	\end{align}
	\end{subequations}
	Then, the fast-fading process corresponding to $T_{c,s}$ and $L_s$ is selected from LUT and assigned to the link $m$.
	
	Hereby, the GSCM and Jake's fading processes are simulated once per given simulation setup and pre-processed to obtain channel statistics of GSCM that are stored in tuple data. This generated tuple data can be then re-used for different simulations and parameters which saves additional computation complexity. Reuse of the pre-processed tuple data is further enhanced by introducing scaling of the data with respect to user velocity and carrier frequency which prevents the repetition of the pre-processing.

	\section{Mobility Models, Simulation Scenario and Parameters}
	\label{sec:SimScenario}
	
	In this section, the investigated scenario, mobility and propagation parameters are described. These will be used to compare the different mobility performance indicators obtained for the simplified channel model against those from Jake's channel model for various simulation cases.
	
	\begin{figure}[!htb]
	\centering
	\includegraphics[width=\columnwidth]{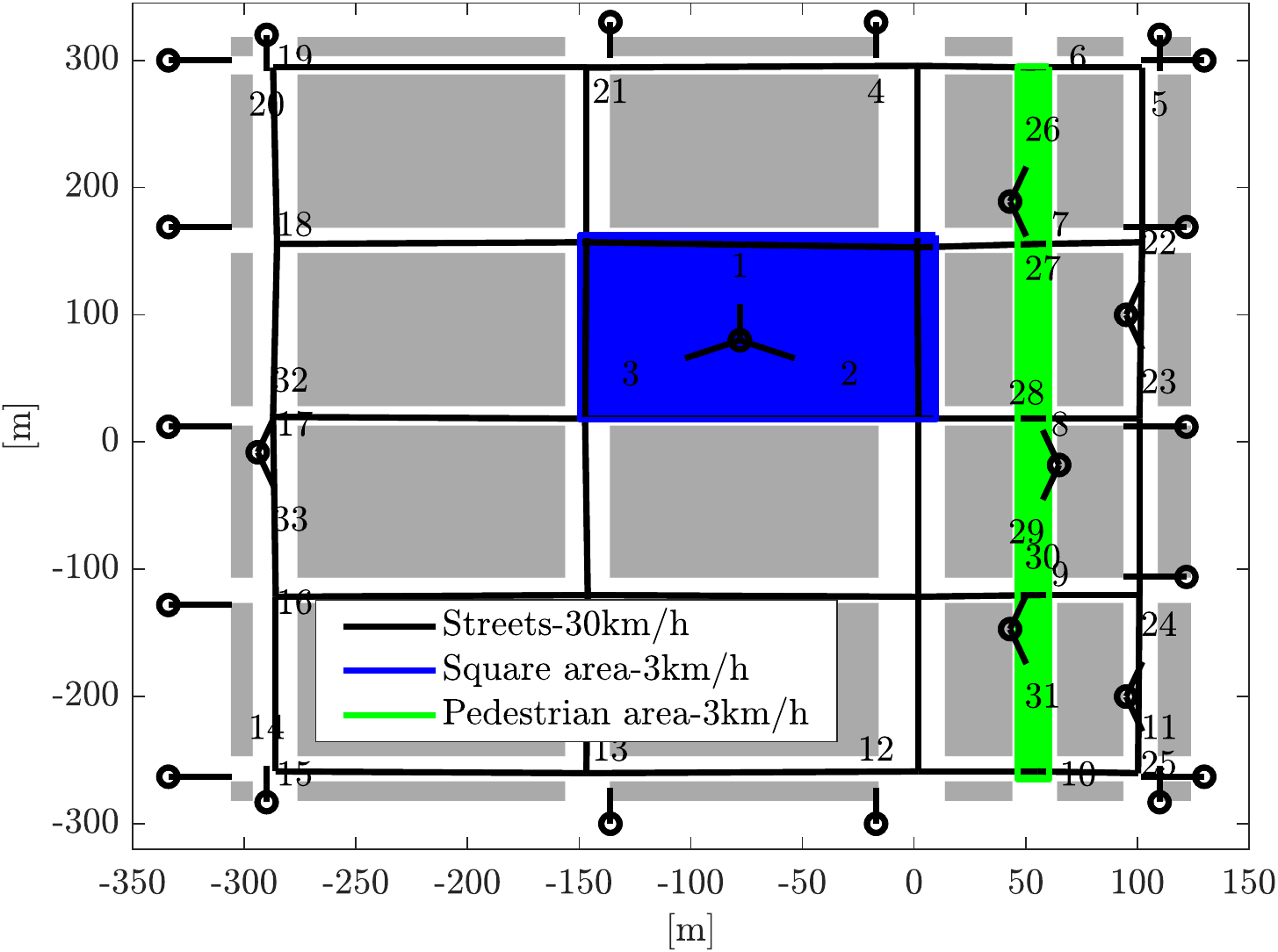}
	\caption{Madrid Grid layout is used for simulation scenarioes that is described in METIS 2 project \cite{METIS2}. The scenario consists of buildings (grey), streets (black) with 200 users, open square (blue) with 40 users and pedestrian area (green) with 80 users.}
	\label{fig:environment}
	\end{figure}

	In this study, the Madrid Grid layout that is described in the METIS 2 project \cite{METIS2} is used. The layout is given in Figure \ref{fig:environment} and consists of buildings (grey), streets (black), open square (blue) and pedestrian area (green). There are $C=33$ 3-sector macro cells which are located on the roof tops of the buildings. The users are distributed as follows: 200 users are moving in the streets with $30$ km/h in both directions. Besides, 40 pedestrian users are walking in the open square and 80 users are walking in the pedestrian area with $3$ km/h.
	
	The scenario parameters along with the configuration of the transmit antenna panels are defined in Table~\ref{tab:sim_parameters1}, Table~\ref{tab:sim_parameters2} and (\ref{eq:beam_ele}). The beams $ b\in[1,8]$ have smaller beamwidth and higher beamforming gain to cover the far regions of the coverage areas whereas beams $ b \in [9,12]$ with larger beamwidth and relatively smaller beamforming gain are defined to serve regions near to the cell. SINR $ \gamma_{c,b}(m) $ of a link between UE and beam $ b $ of cell $ c $ is evaluated by the approximation given in \cite{SINRModel} for strict resource fair scheduler.
	\begin{table}[!htb]
	\renewcommand{\arraystretch}{1.3}
	\caption{Simulation Parameters \rom{2}}
	\label{tab:sim_parameters2}
	\centering
	\begin{tabulary}{\columnwidth}{L L}
		
		\hline 
		\textbf{Parameters} & \textbf{Value}\\ 
		\hline \hline
		Carrier frequency & 28 GHz\\ 
		
		Number of cells $C$ & $33$\\
		
		Network topology & Madrid grid \cite{METIS2} \\
		
		PRB bandwidth & 10 MHz \\
		
		Downlink TX power & 12 dBm/PRB \\
		
		Thermal noise power & -97 dBm/PRB \\ 
		
		Propagation loss & deterministic model of \cite{SimplifiedDeterministic} \\
		
		Penetration loss & 0 dB \\
		
		Total number $U$ of UEs & $320$\\
		
		Number of simultaneously scheduled beams per cell & $1$\\
		
		Cell-pair specific offset $O^{\textrm{A}_{3}}$ & $3$ dB\\
				
	\end{tabulary} 
	\end{table}
	
	\textit{RSRP and measurement error}: 
	Each UE measures and reports the reference signal received power (RSRP) measurements for serving and neighboring cells which are used by the network for handover decisions. The RSRP comprises path-loss, antenna gain, shadowing and fast-fading. 
	
	The limited number of reference symbols available in handover measurement bandwidth for RSRP measurements introduces measurement error (ME). This error is modeled as log-normal distributed with zero mean and $\sigma_\epsilon$ dB standard deviation as defined in \cite{MeasErr}.
	
	\textit{Layer-1 and Layer-3 filtering}: Two levels of filtering are applied to RSRP measurements to mitigate the effects of the fast-fading and measurement errors: Layer-1 (L1) filtering applies moving average to RSRP measurements over a certain bandwidth on the physical layer and outputs $Q(n)$, where $n$ is the discrete time step of the L1 output \cite{Meryem}. 
	
	Layer-3 (L3) filtering further averages the L1 output $Q_{\textrm{dB}}(n)$ by applying an infinite impulse response (IIR) filter in dB \cite{Meryem}. L3 output $\overline{Q}_{\textrm{dB}}(n+1)$ is evaluated as
	\begin{equation}
		\centering
		\label{eq:L3Filter}
		\overline{Q}_{\textrm{dB}}(n+1) = \alpha Q_{\textrm{dB}}(n+1) + (1-\alpha)\overline{Q}_{\textrm{dB}}(n).
	\end{equation}
	In (\ref{eq:L3Filter}), the forgetting factor $\alpha$ controls the impact of older measurements $\overline{Q}_{\textrm{dB}}(n)$ on the current measurement $
	{{Q}_{\textrm{dB}}(n+1)}$ \cite{38331}. To this end, $\alpha$ is configured using filter time constant $T_\alpha$ where the impact of an L1 RSRP measurement  $Q_{\textrm{dB}}(n)$ is halved after $T_\alpha$ period \cite{38331}. 
	
	\textit{Handover}:	A handover for user $u$ is triggered by the serving cell when it receives L3 measurements reported from the UE. This measurement report is sent at time instant $ n = n_0 $ if the following condition expires at the UE,
	
	\begin{equation}
		\centering
		\label{eq:A3}
		\overline{Q}_{\textrm{dB},c_0}(n) + O^{\textrm{A}_{3}}_{c_0,c} < \overline{Q}_{\textrm{dB},c}(n)~~\textrm{for}~~ n_0-T_T<n<n_0,
	\end{equation}
	where $O^{\textrm{A}_{3}}_{c_0,c}$ is a cell-pair specific offset that is configured by the serving cell $ c_0 $ for each neighboring cell $ c $. In (\ref{eq:A3}) the condition expires at $n=n_0$ when the L3 measurement $\overline{Q}_{\textrm{dB},c}(n)$ of a neighbor cell $ c\neq c_0 $ exceeds that of the serving cell power $\overline{Q}_{\textrm{dB},c_0}(n)$ for a certain amount time $T_T$ which is called time to trigger.
	
	After receiving the measurement report from the UE, the serving cell sends a HO command to the UE. The UE receives it successfully if its signal-to-interference-noise ratio (SINR) is above a quality threshold $\gamma_{\textrm{thr,out}}$. The handover execution towards the target cell is completed only if the SINR with respect to the target cell exceeds $\gamma_{\textrm{thr,out}}$ during the random access procedure.
	
	\textit{Radio Link Failure}: Radio link failure (RLF) is one of the most important key metric that is used to evaluate mobility performance. An RLF timer $T_{310} = 600$ ms is started when $\gamma_{\textrm{SINR}}$  falls below $\gamma_{\textrm{thr,out}}$ and RLF is declared if $T_{310}$ expires. During the timer, the UE may recover before detecting RLF if SINR exceeds a second threshold $\gamma_{\textrm{thr,in}}$ which is larger than $\gamma_{\textrm{thr,out}}$. A detailed explanation of the procedure is given in \cite{38331}.
	
	\section{Performance Evaluation}
	\label{sec:PerformanceEvaluation}
	In this section, the proposed channel model is compared against Jake's channel model \cite{Jakes} with multipath diversity given in (\ref{eq:jakes_multipath}). The key performance indicators (KPIs) used for comparison are explained below.
	\subsection{KPIs}

	\textit{Outage Percentage}: Outage is a time period when a UE is not able to receive data due to several reasons. Herein, it is assumed that the data transmission is not possible when $\gamma_{\textrm{SINR}}$ is below $\gamma_{\textrm{thr,out}}$. Besides, if $T_{310}$ expires due to RLF, the UE initiates re-establishment to a cell (any cell including serving cell) and during the re-establishment, the UE cannot receive any data. Successful handover also causes a short outage since during the random access period $ T_{\textrm{HO}} $ of the the UE to a new cell, UE cannot receive any data. Outage is evaluated as follows
	\begin{equation}
	\label{eq:OutagePer}
	\centering
	\textrm{Outage (\%)} = \frac{\sum_{u=1}^{U} \textrm{~Outage~of~UE}~u}{U \times \textrm{Simulated~time}} \times 100.
	\end{equation}
	
	$ N $\tsub{HO}: Denotes the number of successful handovers from a serving cell to a neighboring cell in the network. 
	
	$ N $\tsub{RLF}: Denotes the number of RLFs that are declared in the network.
	
	Outage percentage, $N_{\textrm{HO}}$ and $N_{\textrm{RLF}}$ are normalized to the number of UEs and simulation time as illustrated in the following section.
 	\subsection{Simulation Results}
	
\begin{figure}[!htb]
	\centering
	\subfloat[Number of RLFs.] {\includegraphics[width=0.95\columnwidth]{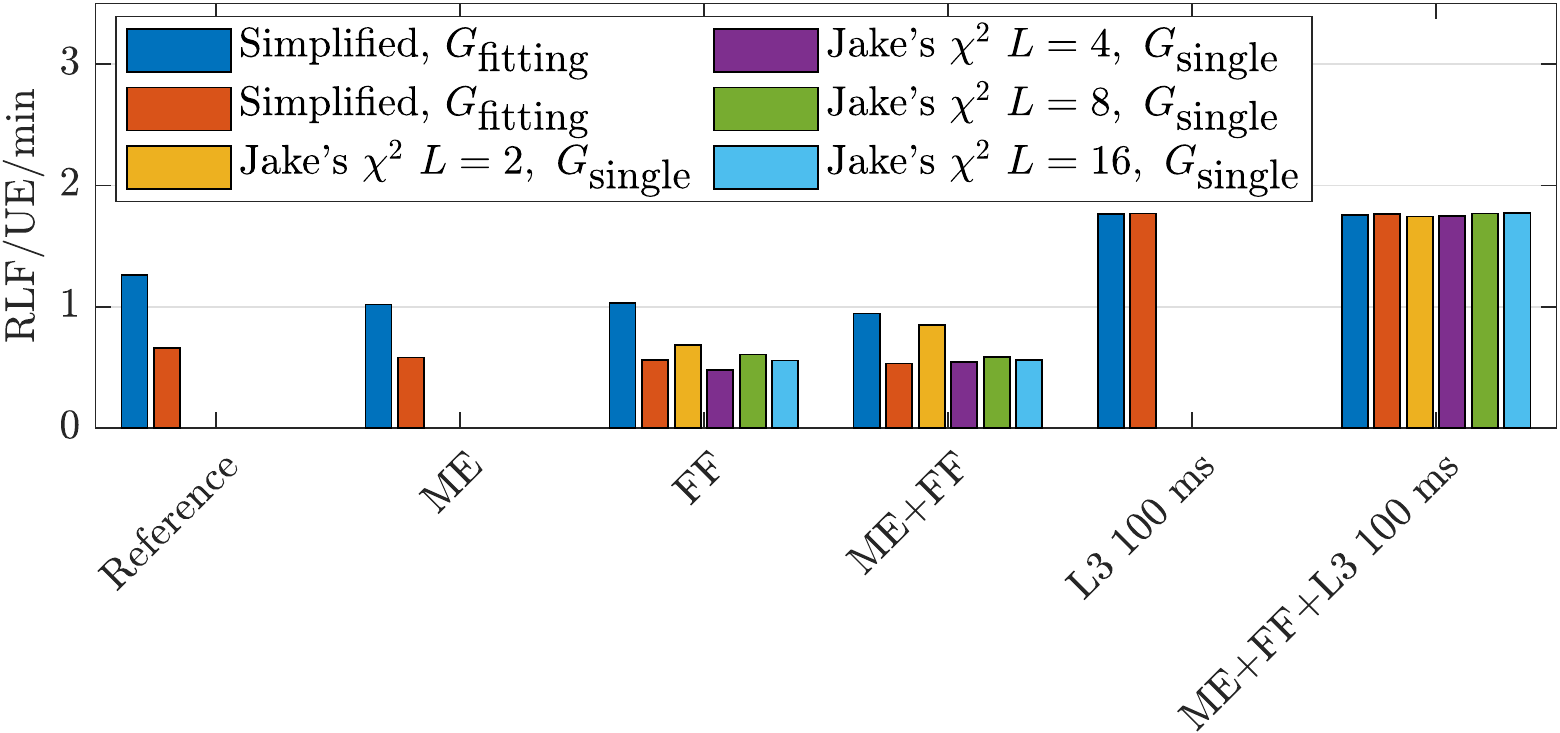}\label{fig:RLF}}
	\hfil
	\subfloat[Number of successful handovers.] {\includegraphics[width=0.95\columnwidth]{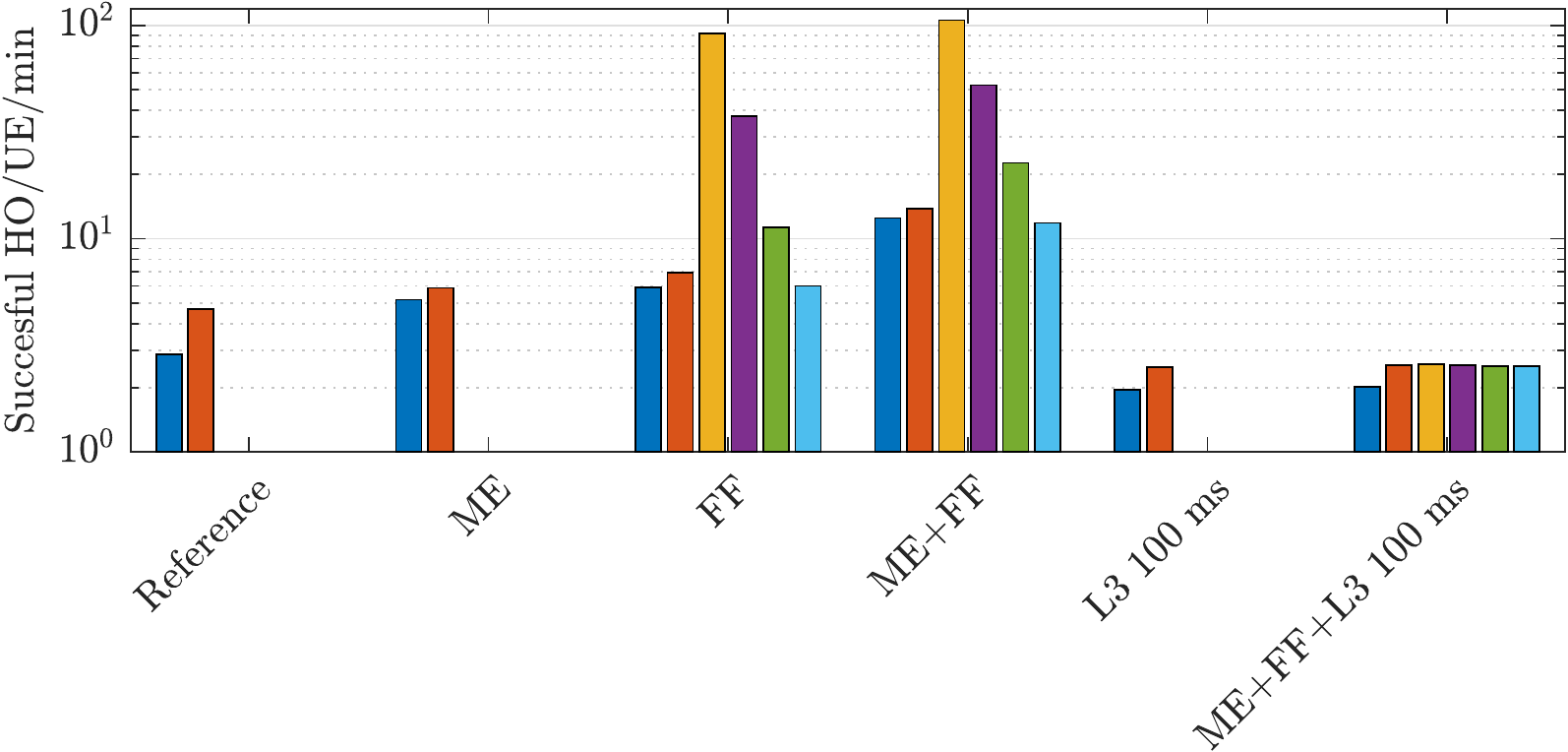}\label{fig:suc_ho}}
	\hfil
	\subfloat[Outage percentage.] {\includegraphics[width=0.95\columnwidth]{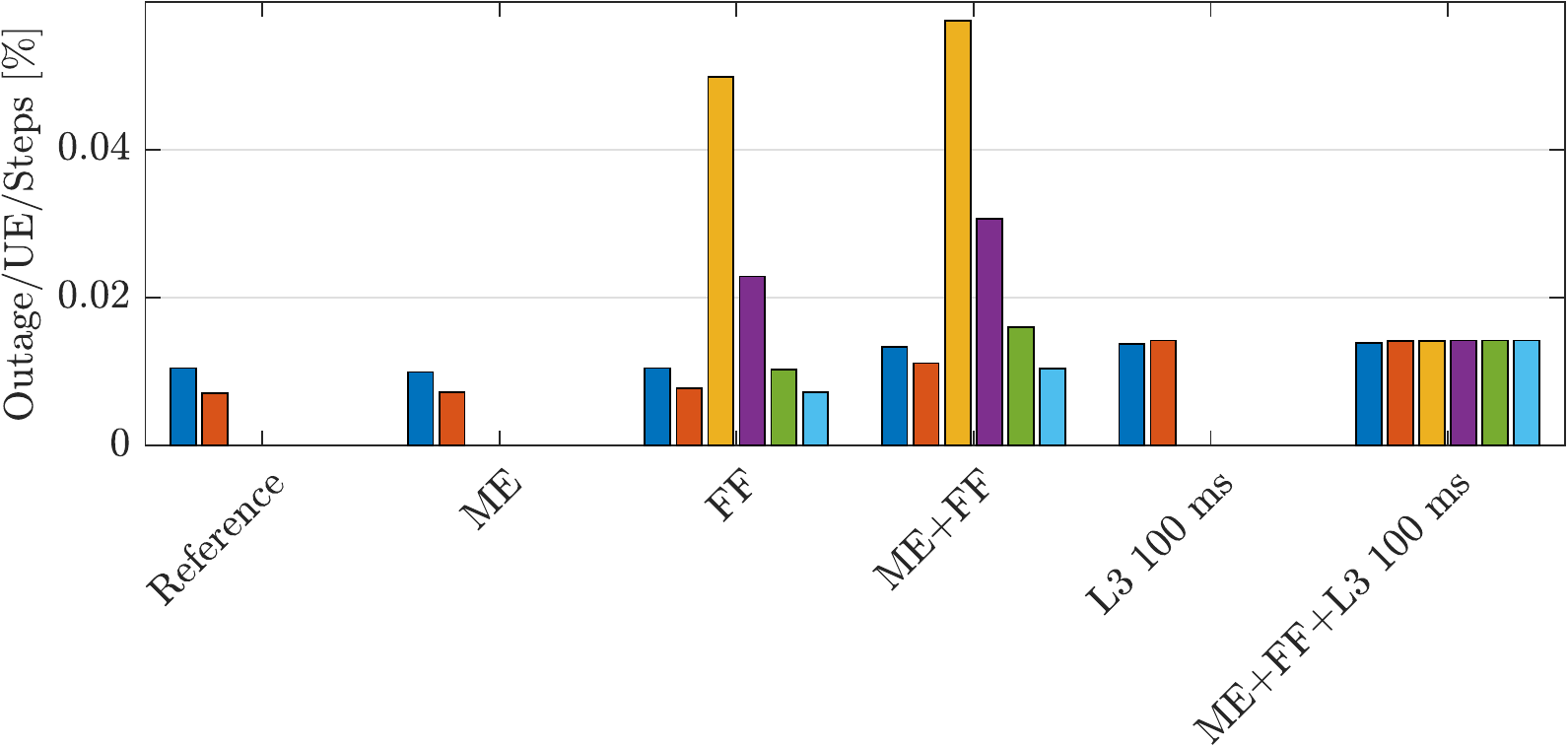}\label{fig:outage}}
	\caption{Impacts of measurement error, fast-fading, L3 filtering on mobility KPIs are shown for simplified and Jake's channel models and for different beamforming gain models. Results are illustrated for different $ L $-path diversity of Jake's channel model.}
	\label{fig:bars}
	\end{figure}

\begin{figure}[!htb]
	\centering
	\subfloat[Number of RLFs.] {\includegraphics[width=0.95\columnwidth]{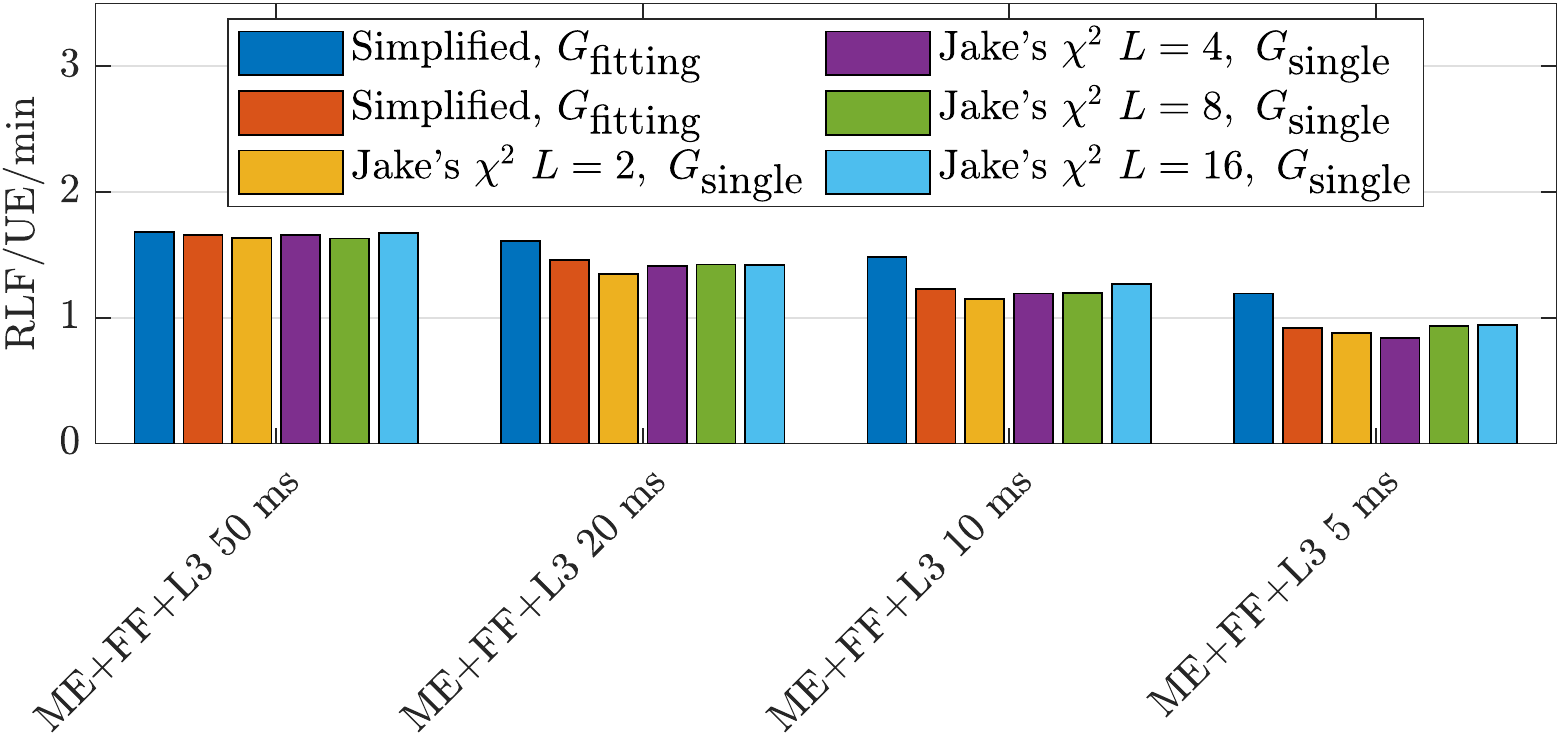}\label{fig:RLF2}}
	\hfil\hfil
	\subfloat[Number of successful handovers.] {\includegraphics[width=0.95\columnwidth]{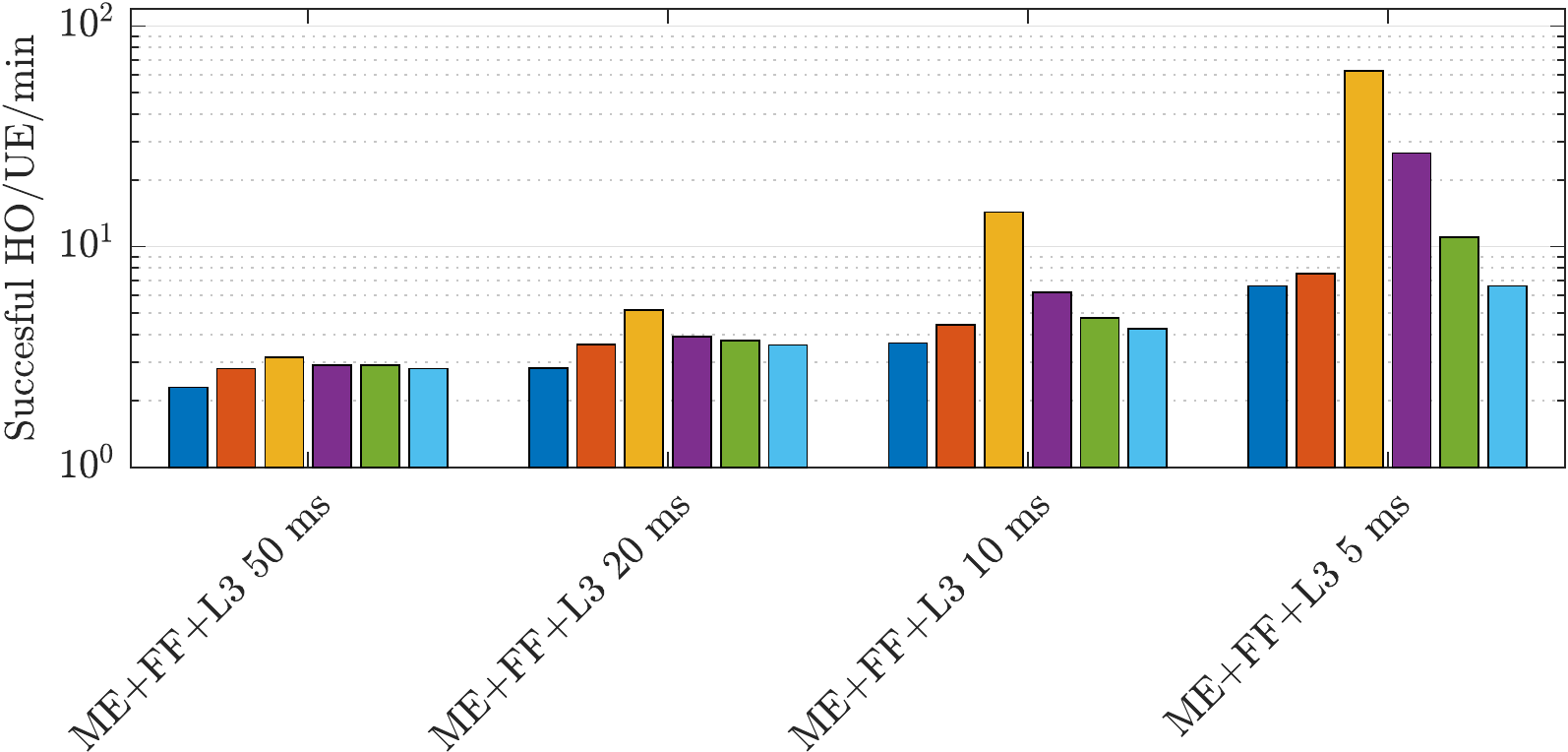}\label{fig:suc_ho2}}	
	\hfil
	\subfloat[Outage percentage.] {\includegraphics[width=0.95\columnwidth]{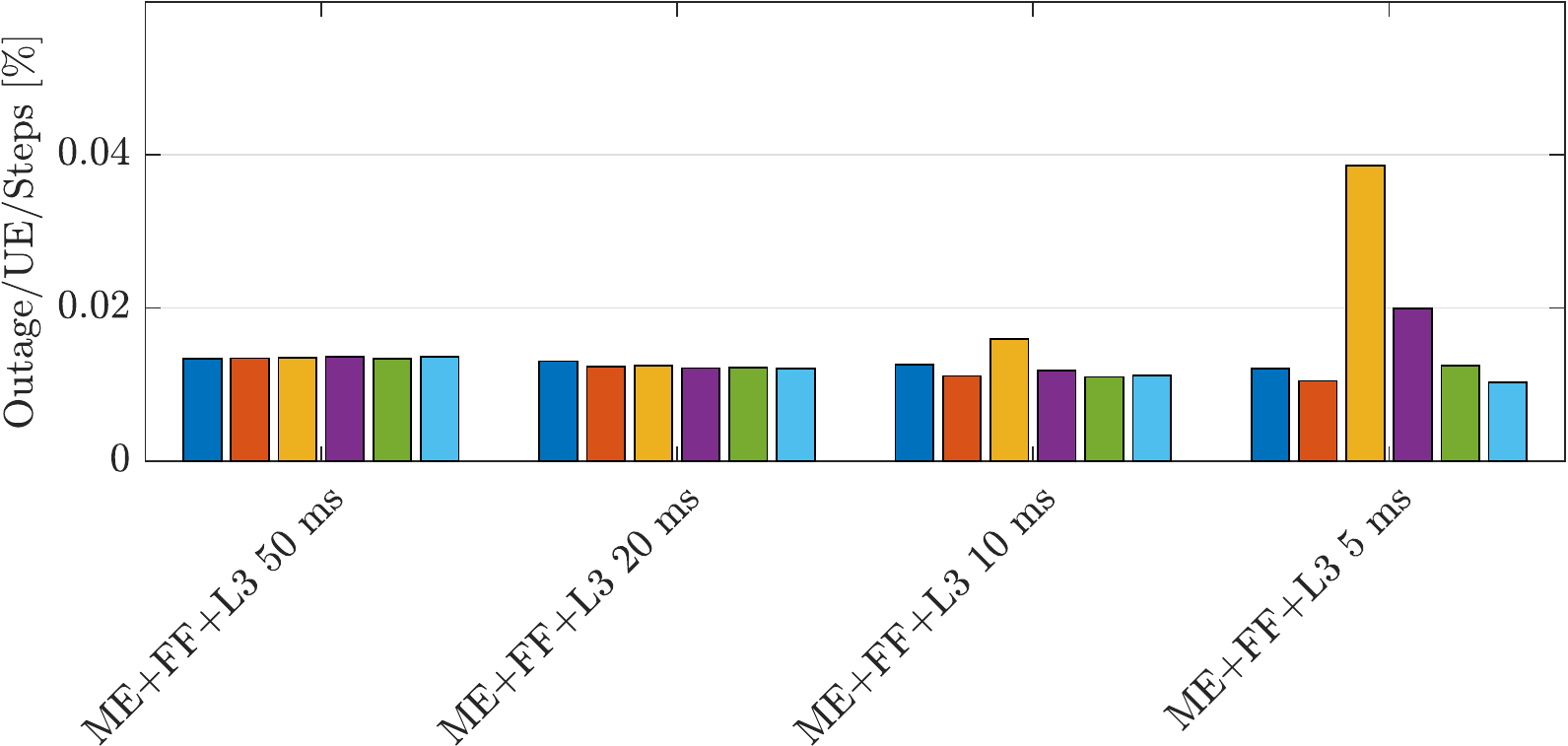}\label{fig:outage2}}
	\caption{Impact of L3 filtering time constant duration on mobility KPIs is shown for simplified and Jake's channel models and for different beamforming gain models. Results are illustrated for different $ L $-path diversity of Jake's channel model.}
	\label{fig:bars2}
	\end{figure}

	The KPIs are shown for different cases (Table~\ref{tab:cases}) of the given scenario setup. In those cases, the impact of beamforming gain models ($ G_{\textrm{single}} $ and $ G_\textrm{fitting} $), measurement error (ME), fast-fading (FF), HO access period $T_\textrm{HO}$ and L3 filtering on the channel model is analyzed. In order to apply $ G_\textrm{fitting} $, $ G_{\textrm{single}} $ is evaluated for each link and the corresponding $ G_\textrm{fitting} $ is obtained from fitting curves (Figure~\ref{fig:BFG_all}) for each LOS and NLOS case. It is assumed that the cell-pair specific offset $O^{\textrm{A}_{3}}$ is set to the same value for each (serving-target) cell-pair. The results are discussed for the proposed simplified and Jake's channel model with $L \in \{2,4,8,16\}$ path diversities. The used beamforming gain models $ G_\textrm{fitting} $ and $ G_\textrm{simple} $ are indicated in the legend of each channel model in Figure~\ref{fig:bars} and Figure~\ref{fig:bars2}.
	
	\begin{table}[!htb]
	\renewcommand{\arraystretch}{1.3}
	\caption{Simulation Case Definitions.}
	\label{tab:cases}
	\centering
	\begin{tabulary}{\columnwidth}{|L||L|L|L|L|}
		\hline 
		\textbf{Simulation Case}&\textbf{FF}&\textbf{ME}&\textbf{L3}&$\boldsymbol{T_\alpha}$  \\ \hline \hline  
		 \textbf{Reference}		&\xmark&\xmark&\xmark&\xmark	\\ \hline
		 \textbf{ME}			&\xmark&\cmark&\xmark&\xmark	\\ \hline
		 \textbf{FF}			&\cmark&\xmark&\xmark&\xmark	\\ \hline
		 \textbf{ME+FF}			&\cmark&\cmark&\xmark&\xmark	\\ \hline
		 \textbf{L3}			&\xmark&\xmark&\cmark&$100$	ms\\ \hline
		 \textbf{ME+FF+L3 100 ms}&\cmark&\cmark&\cmark&$100$	ms\\ \hline
		 \textbf{ME+FF+L3 50 ms}	&\cmark&\cmark&\cmark&$50$ ms\\ \hline
		 \textbf{ME+FF+L3 20 ms}	&\cmark&\cmark&\cmark&$20$ ms\\ \hline
		 \textbf{ME+FF+L3 10 ms}	&\cmark&\cmark&\cmark&$10$ ms\\ \hline
		 \textbf{ME+FF+L3 5 ms}	&\cmark&\cmark&\cmark&$5$ ms\\ \hline
	\end{tabulary} 
	\end{table}
	
	\textit{Impact of beamforming gain}: Figure \ref{fig:RLF} shows that $N_\textrm{RLF}$ for our simplified channel model with $ G_\textrm{fitting} $ is higher than that of the simplified channel model with $ G_\textrm{single} $ and Jake's channel model. The reason for this is that the interference signal originates from non-serving beams, which are mostly directed towards other directions than the serving beam and $G_\textrm{single}$ of non-serving beams mostly yields very low values. In case $ G_\textrm{fitting} $ is used, lower $G_\textrm{single}$ values saturate around $-20$ dB for the LOS case and $0$ dB for the NLOS case, which leads to higher interference power and lower $\gamma_{\textrm{SINR}}$ value. This is also reflected in the $ N_\textrm{HO} $ results (Figure \ref{fig:suc_ho}), where the handovers are not completed due to insufficient $\gamma_{\textrm{SINR}}$ after the measurement report was triggered by the UE.
	
	\textit{Impact of measurement error and fast-fading}: Measurement error and fast-fading cause rapid fluctuations in received signal power, which may lead to unnecessary triggering of handovers. Figure \ref{fig:suc_ho} shows that the number of handovers $ N_\textrm{HO} $ increases compared with the reference case when measurement error and fast-fading components of the channel are enabled in ME and FF cases. Besides, fast-fading causes a larger increase in $ N_\textrm{HO} $ compared with the measurement error since fluctuations on received signal power caused by fast-fading are stronger than those of the measurement error. $ N_\textrm{RLF} $ slightly decreases when measurement error and fast-fading are enabled because RLF timer $T_{\textrm{310}}$ is reset when the signal power exceeds $\gamma_{\textrm{thr,out}}$ which is observed more often due to the rapid fluctuations in signal power.
	
	\textit{Impact of path diversity}: In Figure \ref{fig:suc_ho}, in case of FF and FF+ME results, $ N_\textrm{HO} $ decreases for increasing path-diversity of Jake's model. This is because rapid fluctuations in received signal power become smoother when the path diversity increases which prevents ping-pong effects. For the same cases, $ N_\textrm{HO} $ of the simplified channel model is close to the one of Jake's channel model with $L=16$ path diversity. This is expected because as it is seen in Figure \ref{fig:PerBeamDoF}, the path diversity distribution of our simplified channel model leads to higher path diversity values.
	
	\textit{Impact of L3 filtering}: L3 filtering causes delay for the L3 measurement reported by the UE to the serving cell. In return, fluctuations caused by fast-fading are filtered and smoother received signal power is obtained. When the reference is compared to the L3 case in Figure \ref{fig:suc_ho}, delayed measurements lead to less successful HOs because they cause late HO decisions and the UEs cannot send the measurement report on time due to low $\gamma_{\textrm{SINR}}$. This is also visible in Figure \ref{fig:RLF}, where UEs suffer from low $ \gamma_{\textrm{SINR}} $ and the $ N_\textrm{RLF} $ of case L3 increases significantly compared with the reference case. When L3 is combined with ME and FF (ME+FF+L3), $ N_\textrm{HO} $, $ N_\textrm{RLF} $ and outage results do not change with respect to the L3 case since the impairments caused by measurement error and fast-fading are filtered out but the impact of delayed measurement remains.
	
	\textit{Impact of filter time constant}: Reducing the L3 filter time constant $ T_\alpha $ decreases the latency on the measurements caused by L3 filtering. In Figure \ref{fig:bars2} simulation results of the ME+FF+L3 case are shown for different L3 filtering time constants $T_\alpha$. As a consequence of decreasing $T_\alpha$, the number of successful HO increases for all channel models (Figure \ref{fig:suc_ho2}) because channel impairments are not filtered too much for shorter $T_\alpha$. This impact is more apparent for Jake's channel model with $L=2$ path diversity since the received signal fluctuations are higher for lower path diversities. On the other hand, increasing $T_\alpha$ results in smaller latency of the measurements. Consequently, $ N_\textrm{RLF} $ decreases for increasing $T_\alpha$ since the HO is triggered before $ \gamma_{\textrm{SINR}} $ drops below $\gamma_{\textrm{thr,out}}$.
	
	\section{Conclusion}
	\label{sec:Conclusion}
	
	In this paper, a simplified channel model is presented for beamforming systems operating at mmWave frequencies considering the spatial and temporal characteristics of the propagation channel. Stochastic properties of the channel model developed by 3GPP are studied and those properties are reflected in the simplified channel model by adopting Jake's channel model. Furthermore, we show that the geometry-based beamforming gain is correlated with the beamforming gain which is obtained by applying a directional transmission scheme to the 3GPP channel model. The impact of the channel model on mobility key performance indicators such as successful handover, radio link failure and outage is investigated for different propagation channel configurations. In addition, the influence of measurement error, fast-fading and L3 filtering is analyzed. Results have shown that the proposed channel model provides close simulation results to Jake's model for high path diversity. Moreover, it is shown that the angular spread of the rays impacts the beamforming gains of interfering beams which leads to smaller SINRs and increased RLFs. Results also show that the simplified channel model and Jake's model with high path diversity have similar mobility performance. However, this similarity cannot be generalized as path diversity order of Jake’s channel model may need to be determined for each mobility scenario by comparing, for instance, the mobility results with those obtained from the simplified channel model.
	
	\bibliographystyle{IEEEtran}
	\bibliography{references}

\end{document}